\documentclass[useAMS]{mn2e}

\usepackage{graphicx,amssymb}

\title[Resolved photometry of YMCs in NGC 4214]{Resolved photometry of Young Massive Clusters in the
starburst galaxy NGC 4214\thanks{Based on observations made with the NASA/ESA
Hubble Space Telescope, which is operated by the association of Universities for
Research in Astronomy, Inc., under the NASA contract NAS 5-26555, under program
GO-10332 (PI: Ford)}}
\author[Sollima et al.]{A. Sollima$^{1}$\thanks{E-mail:
antonio.sollima@oabo.inaf.it}, M. Cignoni$^{1,2}$, R. G. Gratton$^{3}$, M.
Tosi$^{1}$, A. Bragaglia$^{1}$,
\newauthor S. Lucatello$^{3}$, G. Meurer$^{4}$\\
$^{1}$ INAF Osservatorio Astronomico di Bologna, via Ranzani 1, Bologna, 40127,
Italy\\
$^{2}$ Dipartimento di Astronomia, Universit\'a di Bologna, via Ranzani 1, Bologna, 40127,
Italy\\
$^{3}$ INAF Osservatorio Astronomico di Padova, vicolo dell'Osservatorio 5,
Padova, 35122, Italy\\
$^{4}$ International Centre for Radio Astronomy Research, The University of Western 
Australia, 35 Stirling Highway, Crawley, WA 6009, Australia}
\begin{document}


\pagerange{\pageref{firstpage}--\pageref{lastpage}} \pubyear{2013}

\maketitle

\label{firstpage}

\begin{abstract}
  We present the results of deep high resolution imaging performed
  with ACS/HRC@HST in the most active region of the nearby starburst
  galaxy NGC 4214. We resolved the stellar populations of five Young
  Massive Clusters and their surrounding galactic field. The star
  formation history of this region is characterized by two main bursts
  occurred within the last 500 Myr, with the oldest episode spread
  out across an area larger than that covered by the most recent
  one. The ages derived for the analysed clusters cover a wide range
  between $6.4<\log{t/yr}<8.1$ in agreement with those predicted by
  recent analyses based on integrated photometry. The comparison between
  the mass of the young associations and that of the surrounding field
  population with similar ages indicates a high cluster formation efficiency 
  ($\Gamma\sim33$\%) which decreases when old populations are considered.
  The mass function
  of the major assembly has been found to be slightly flatter than the
  Salpeter (1955) law with a hint of mass segregation. We found no
  clear signatures of multiple stellar populations in the two young
  ($\log{t/yr}<6.8$) associations where we were able to resolve their
  innermost region. The masses and sizes of three
  clusters indicate that at least one of them could evolve toward a
  globular cluster-like structure.
\end{abstract}


\section{Introduction}
\label{intro_sec}

The study of the star formation histories (SFH) of dwarf galaxies and the
relation with their population of massive clusters constitutes a crucial piece 
of information to understand how star formation proceeded in the early stages of
formation of structures in the Universe. 
One of the main cosmological interests is related to the possibility that 
today's dwarfs are the remnants of the building blocks of massive galaxies. 
In the context of the widely accepted cosmological model $\Lambda$-Cold Dark Matter, 
more massive systems are assembled by subsequent merging of 
these protogalactic fragments (the hierarchical formation scenario; e.g., White
\& Rees 1978; Frenk et al. 1988).
The SFH of dwarfs with different morphological
type are quite different (see Tolstoy, Hill \& Tosi 2009 for a recent
review). In particular, early-type dwarfs have little or no SF in the 
last few Gyrs (e.g. Weisz et al 2011), whilst all late-type dwarfs experience 
more recent star formation
activities, with major peaks in the last few Gyrs (e.g. Cole et al 2007; 
Cignoni et al 2013), although many of them appear to have formed
the majority of their stellar mass prior to 7.6 Gyr ago (Weisz et al 2011).
Among the latter group, starburst galaxies, characterized by a recent intense SF
activity, constitute an interesting class of objects. Although they are quite
rare in the local Universe, their fraction increases at high redshift (Vieira et
al. 2013) indicating a larger frequency of starbursts in the early Universe.
As all the starburst galaxies in the nearby Universe are located outside the
Local Group, their SFH could be determined only in recent years through Hubble
Space Telescope observations (Greggio et al. 1998; Aloisi, Tosi \& Greggio 1999; Silva-Villa \& Larsen 2012; 
Garc{\'{\i}}a-Benito \& P{\'e}rez-Montero 2012).

Massive clusters are important
tracers of the stellar content of their host galaxy and their study provides
crucial insight on the understanding of the star formation process at different
epochs. For instance, the correlation of the properties (colours, magnitudes) of old 
globular clusters (GCs) with those of their host galaxy is a valuable 
information to understand whether its GC system formed in-situ or was accreted
(Larsen et al. 2001). On the other hand, young clusters provide information on
the recent star formation activity of galaxies.
It has been suggested that up to $\sim$40\% of the stars in galaxies formed in
temporary bound systems (Kruijssen 2012) which later dissolve as an effect of the 
expulsion of residual gas by feedback (Tutukov 1978; Parmentier et al. 2008). 
Young Massive Clusters (YMCs; with $\log{t/yr}<8$ and
$\log{M/M_{\odot}}>4$, Portegies-Zwart, McMillan \& Gieles 2010) are
among the most interesting objects in this context. They have
been observed in the Milky Way and the Local Group galaxies, but their
fraction is strikingly high in starburst and interacting galaxies,
being tightly correlated with the star formation rate of the host
galaxy (Larsen \& Richtler 2000; Billett, Hunter \& Elmegreen 2002).
YMCs trace the recent star forming activity of distant galaxies and
may be used to study the star-cluster mass function. 
On the other
hand, the closest YMCs are of prime interest to study the shape and
the homogeneity of the stellar mass function in an environment which
has still not been affected by dynamical processes like two-body
relaxation, and the interplay between stellar evolution and stellar
dynamics. Furthermore, they are the best candidates to be the
progenitors of GCs, sharing in some cases similar masses
and sizes. Their young ages allow therefore to study the early
evolution of massive star clusters in the stage when they are expected
to form multiple generation of stars (e.g. D'Ercole et al. 2008). While the
integrated properties of YMCs have been studied in detail in the past
decades (Arp \& Sandage 1985; Silva-Villa \& Larsen 2011), resolved
photometry has been feasible only for Milky Way YMCs (Figer et
al. 1999, 2006; Davies et al. 2007) and, only very recently, in nearby
galaxies (Perina et al. 2010; Larsen et al. 2011).

NGC 4214 is a nearby dwarf IAB(s)m galaxy (de Vaucouleurs et al. 1991)
in the low-redshift CVn I Cloud (Sandage \& Bedke 1994) located at a
distance of $3.04 \pm 0.05$ Mpc (Dalcanton et al. 2009). It has a mass
of $M\sim5\times 10^{9}~M_{\odot}$ (Karachentsev et al. 2004) and a
low metal-content ($-1.6<[M/H]<0$; Williams et al. 2011
hereafter W11). NGC 4214 is characterized by an intense recent
star-formation activity, as shown by the presence of two HII star
forming complexes located in its central region (Fanelli et al. 1997).
Recently, McQuinn et al. (2010) and W11 used Hubble Space Telescope (HST) 
photometric
observations to study the field population of NGC4214. They found a SFH
characterized by an old episode occurred more than 2 Gyr ago followed by an
intense recent SF activity.
The large majority of the galaxy stellar population ($\sim 75\%$ of
the total galaxy mass) appears to be
homogeneously distributed across the galaxy except for the young stellar
population which is more centrally concentrated. MacKenty et al. (2000)
provided a detailed classification of the $H\alpha$ knots visible in
HST images, identifying 13 relatively bright YMCs.
Recently, {\'U}beda,
Ma{\'{\i}}z-Apell{\'a}niz \& MacKenty et al. (2007a,b) and Sollima et
al. (2013) used imaging from the WFPC2@HST associated to infrared
photometry to determine ages, masses and extinction of a sample of
YMCs. In both the above studies YMCs cover a wide range in age
($6.2<\log{t/yr}<8.3$) and masses ($3.7<\log{M/M_{\odot}}<6$).  Dust
absorption appears to be on average relatively low ($E(B-V)<0.1$; W11,
Sollima et al. 2013) although patchy with some regions heavily
obscured (Ma{\'{\i}}z-Apell{\'a}niz 1998; Beck, Turner \& Kovo 2000;
{\'U}beda et al. 2007a,b). 

In this paper we present the resolved photometry of five young stellar complexes
of NGC 4214 obtained using ACS@HST. These data are used to determine ages, masses
and half-light radii of these objects and the SFH of
their surrounding field. Throughout the paper we use the naming
convention for clusters by Ubeda et al. (2007a; see their Fig. 2). 
The paper is organized as follows: in
Sect. 2 we describe the observations and the data reduction
technique. In Sect. 3 the obtained colour-magnitude diagrams (CMDs) are
presented. In Sect. 4 the SFH of the field population is presented and
discussed.  Sect. 5 is devoted to the determination of ages, masses
and half-light radii of the YMCs and associations of NGC 4214 and to the determination
of the mass function and the degree of primordial mass segregation of
the resolved association I-A. Finally, we summarize and discuss our
results in Sect. 6.

\section{Observations and Data Reduction} 
\label{obs_sec}

\begin{figure*}
 \includegraphics[width=12cm]{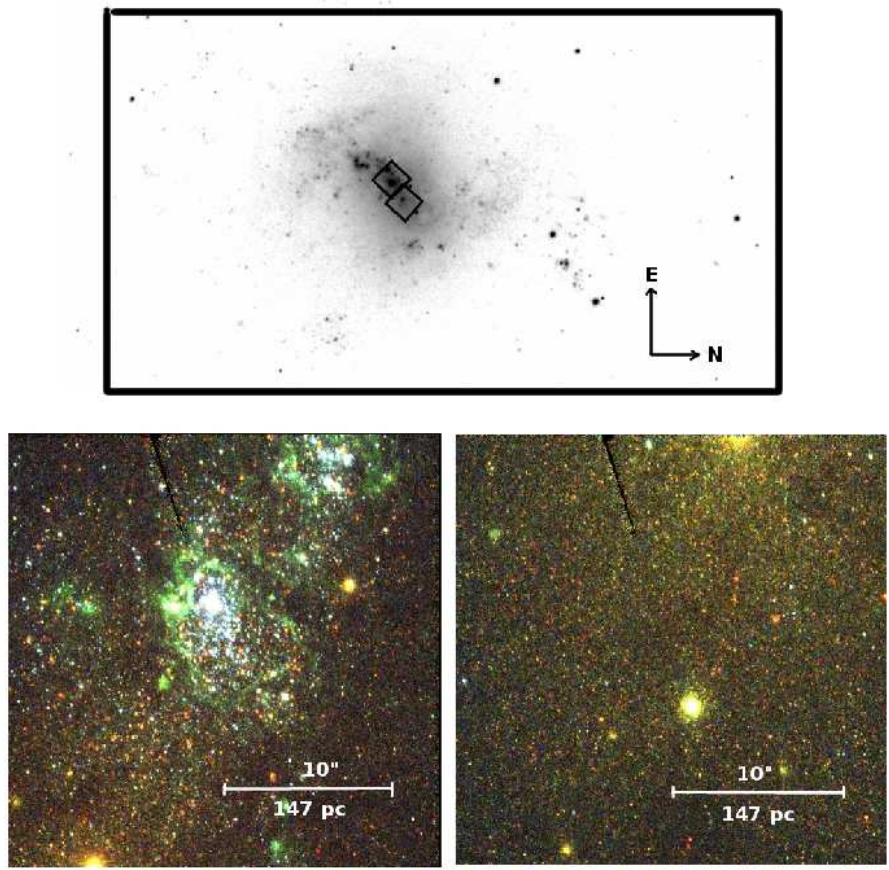}
 \caption{Location of the two HRC/ACS pointings on the SDSS $g$ image of
   NGC4214 (top panel). In the lower panels the false-colour image of the
   two HRC pointings of NGC 4214 (left: central field; right: offset
   field) are shown. The F330W, F555W and F814W images have been
   combined in the blue, green and red channels, respectively (in the
   printed version of the paper the F555W band image is shown in
   greyscale). The angular and physical scale (adopting a distance of
   3.04 Mpc) are indicated in both panels.}
\label{ima}
\end{figure*}

The analysed dataset consists of a set of images obtained with the
High Resolution Channel (HRC) of the Advanced Camera for Survey onboard HST
under the program GO-10332. This camera provides a field of view of
$29\arcsec\times26\arcsec$ with a plate scale of
0.027$\arcsec~px^{-1}$. Two regions of the main star forming complex
4214-I have been observed through the F330W, F555W and F814W filters:
{\it i)} a pointing centered on the largest association (I-A) and
containing also cluster I-Es, and a portion of the association I-B (hereafter
referred as {\it central field}), and {\it ii)} a pointing centered on
cluster IVs which includes also cluster IIIs at the edge of the frame
(hereafter {\it offset field}). The two fields are offset by
$\sim30 \arcsec$ ($\sim$ 440 pc at the distance of NGC 4214) with only
a marginal overlap. The false colour images of the two pointings are
shown in Fig. \ref{ima}. In the {\it central field} the two main
associations I-A and I-B are evident as two distinct assembly of blue
objects immersed in a population of red stars. Two diffuse shells
surrounding the two main associations are particularly evident in the
F555W image as a consequence of the strong $H\beta$ and $[OIII]$ emission of
gas. The {\it offset field} is instead dominated by red stars covering
the entire field of view.  Three exposures for each filter have been
obtained with exposure times of 223, 130 and 120 s in the F330W, F555W
and F814W filters, respectively. All images were passed through the
CALACS reduction pipeline. Data reduction has been performed
on the individual pre-reduced (.flt) images using the {\rm DAOPHOT II}
package (Stetson 1987). For each image an empirical Point Spread Function has
been determined using $\sim$50 isolated bright stars. Source detection has been performed on the
stack of all images while the photometric analysis was performed
independently on each image. Only stars detected in at least two out of three
long exposures have been included in the final catalog. The final
magnitudes have been obtained as the average of single exposure
measures and the related r.m.s. has been assigned as their error. We
used the most isolated and brightest stars in the field to link the
aperture magnitudes at 0.5 arcsec to the instrumental ones, after
normalizing for exposure time. Instrumental magnitudes have been then
transformed into the VEGAMAG system using the photometric zero
points by Sirianni et al. (2005). Our photometry has been compared with the
catalog by Ubeda et al. (2007a) obtained from WFPC2 data in the F336W and F555W
filters. The mean magnitude differences are
$F336W_{WPFC2}-F330W_{HRC}=-0.20\pm0.05$ and
$F555W_{WFPC2}-F555W_{HRC}=-0.05\pm0.03$. The above differences can be entirely
addressed to the different filters adopted by these authors (see Sirianni et al.
2005).

We performed artificial star
experiments on the science frames: a set of artificial stars have been simulated 
using the same PSF model extracted in the science frames and added to all
images at random positions within 24$\times$24 pixel cells centered on
a grid of 20$\times$20 positions along the x and y directions of the chip
(a single star for each cell). The magnitudes of the artificial stars have been
assigned following a homogeneous distribution in F555W magnitude and
(F330W-F555W) and (F555W-F814W) colours within the ranges
$18.5<F555W<26.5$, $-2<(F330W-F555W)<1$ and $-1<(F555W-F814W)<3$.  We
performed the photometric reduction on the simulated frames with the
same procedure adopted for the science frames producing a catalog of
$\sim$200,000 artificial stars which has been used to determine the
level of completeness and photometric errors in different positions of
the frame.

\section{Colour-magnitude diagrams}
\label{cmd_sec}

\begin{figure*}
 \includegraphics[width=12cm]{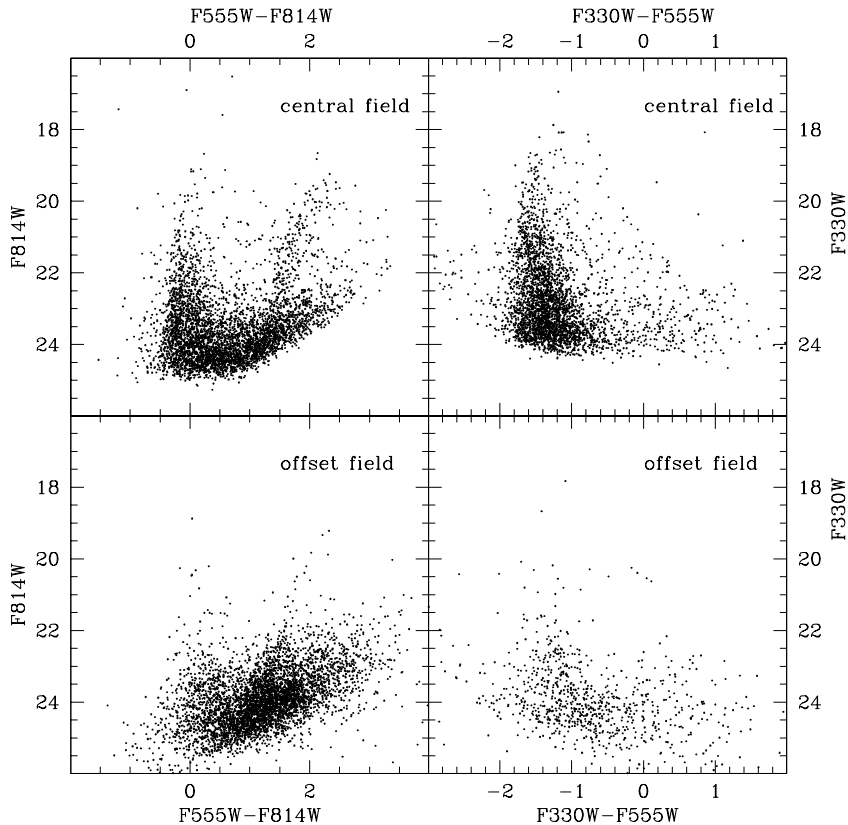}
 \caption{CMDs of the two ACS pointings of NGC 4214 (top: {\it central
     field}; bottom: {\it offset field}). Left panels show the
   F814W vs. (F555W-F814W) CMDs, right panels show the (F330W-F555W)
   vs. F330W CMDs.}
\label{cmd}
\end{figure*}

In Fig. \ref{cmd} the CMDs of the two observed fields are shown in the
F814W vs. (F555W-F814W) and F330W vs. (F330W-F555W) planes. The two
diagrams sample the evolved population of NGC 4214 down to
F555W$\sim$26, just above the tip of the Red Giant Branch (see
W11). The optical CMDs of the two fields appear significantly
different from each other. In fact, in the {\it central field} the prominent blue
plume (BP), populated by massive ($M>10~M_{\odot}$) main sequence
(MS) stars and blue Helium-burning (HeB) stars in the range
$6<M<16~M_{\odot}$, is well visible in the blue region of the CMD
at $(F555W-F814W)<1$. At red colours, the population of Red Supergiants
(RSG; mainly red intermediate to massive HeB stars), form a well
defined sequence at $(F555W-F814W)>1$ and $F814W<21.5$. 
At fainter magnitudes, the red part of this CMD is populated by a
significant fraction of intermediate mass ($M\sim3~M_{\odot}$)
Asympthotic Giant Branch (AGB) stars. In the
optical CMD of the {\it offset field} the fraction of both Blue and
Red Supergiants is significantly smaller than in the {\it central
  field}: only a bunch of blue MS/HeB stars is visible, while the RSG
sequence still populates the bright red portion of the CMD.

In the F330W vs. (F330W-F555W) CMDs most of the stars are
intermediate to massive MS objects with color around -1.5, while the
few objects at the right of the MS are likely HeB stars. 
It is apparent that the recent star formation
  (SF) activity of the {\it central field} is more intense than that
in the {\it offset field}.

\section{The field population}

\subsection{Star Formation History}
\label{sfh_sec}

\begin{figure*}
 \includegraphics[width=13.8cm]{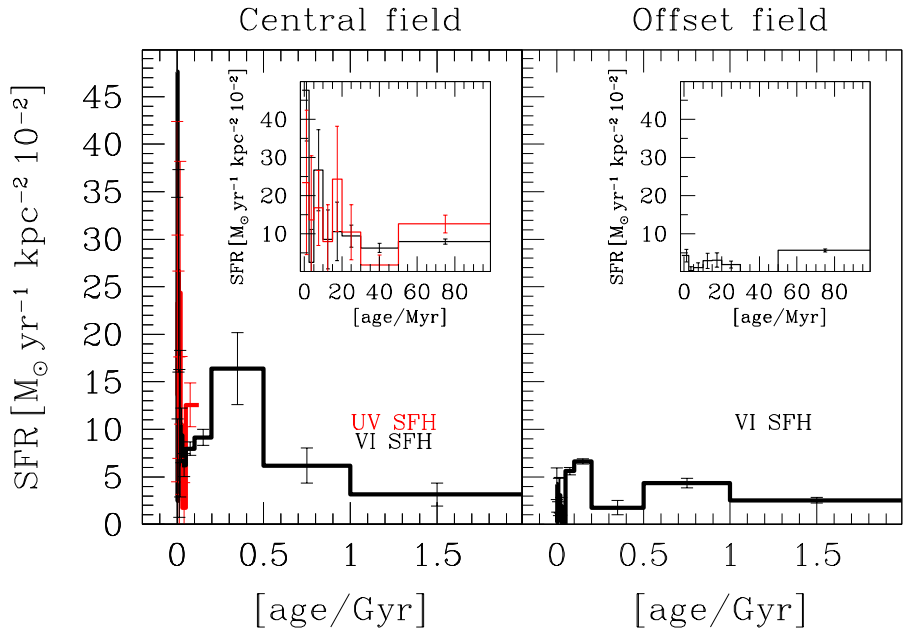}
 \caption{SFH derived for the {\it central field} (left panel) and the
   {\it offset field} (right panel). The black histograms show the SFH
   derived from the F555W vs. (F555W-F814W) CMDs, the red histograms (grey in the
   printed version of the paper) in
   the left panel show the SFH derived from the F555W vs. (F330W-F555W)
   CMD. The inner boxes show a zoom in of the first 100 Myr.}
\label{sfh_field}
\end{figure*}

\begin{figure*}
 \includegraphics[width=12cm]{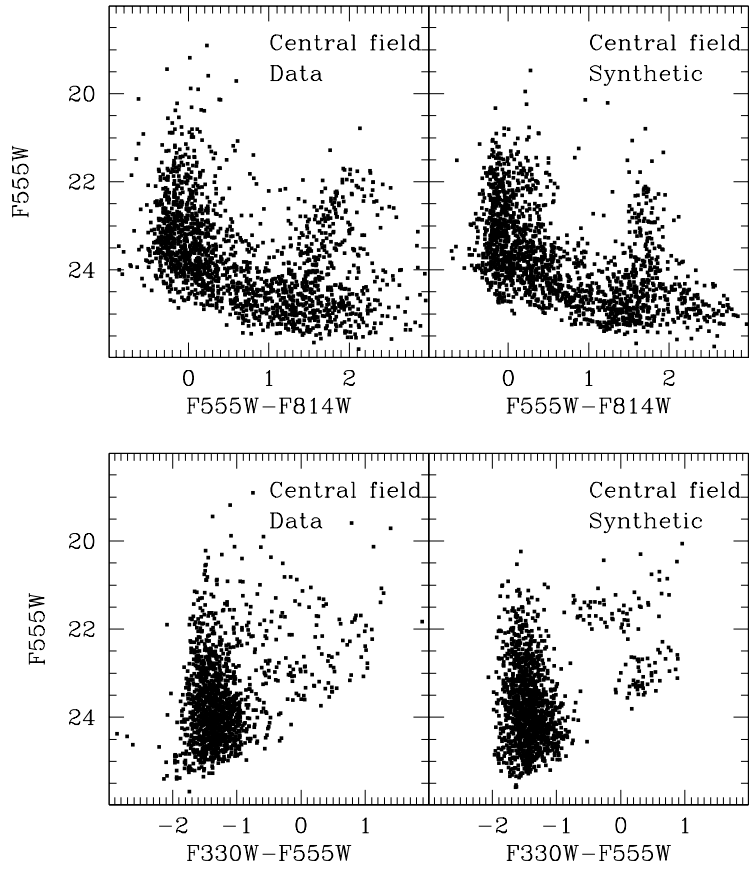}
 \caption{Comparison between the observed (left panels) and synthetic
   (right panels) CMDs of the {\it central field}. Top panels show the
   F555W vs. (F555W-F814W) CMD, bottom panels show the 
   F555W vs. (F330W-F555W) CMD.}
\label{cmd_field1}
\end{figure*}

\begin{figure*}
 \includegraphics[width=12cm]{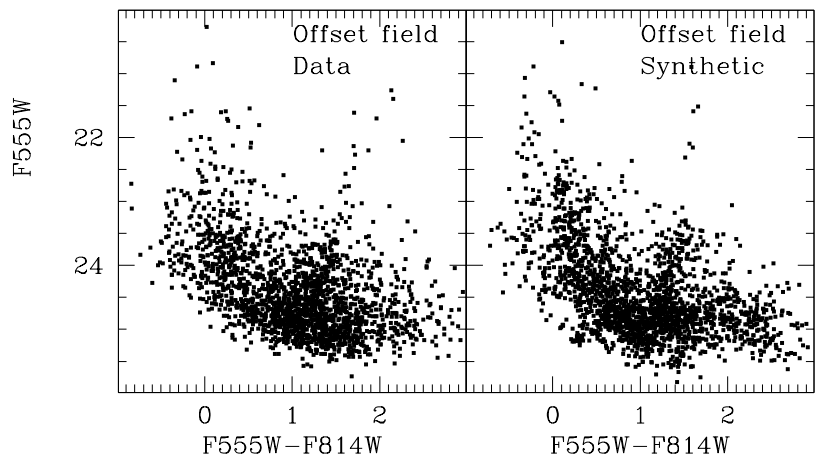}
 \caption{Same of Fig.\ref{cmd_field1} but for the {\it offset
     field}. Only the F555W vs. (F555W-F814W) CMD has been used for this
   field to derive the SFH.}
\label{cmd_field2}
\end{figure*}

\begin{figure*}
 \includegraphics[width=12cm]{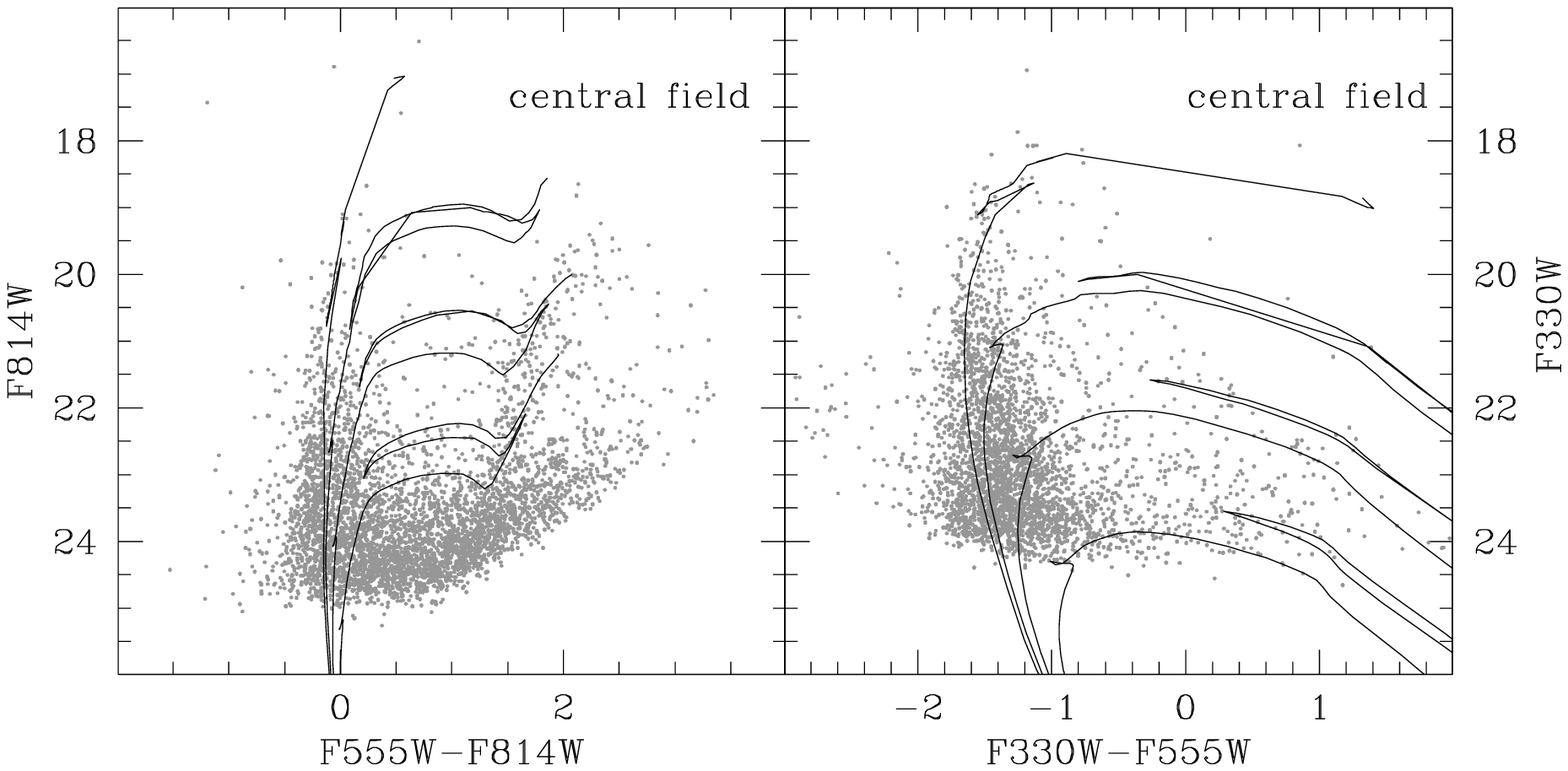}
 \caption{CMDs of the {\it central
     field}. The left panel shows the
   F814W vs. (F555W-F814W) CMD, the right panel shows the 
   F330W vs. (F330W-F555W) CMD. A set of Marigo et al. (2008) isochrones with Z=0.008 and
   $\log{t/yr}$=6.6, 7, 7.4 and 7.8 are overplotted in
   both panels.}
\label{cmdiso}
\end{figure*}

We use the CMDs shown in Fig. \ref{cmd} to determine the star formation history
(SFH) of the
field population surrounding the YMCs and associations in the two observed regions. We
adopted the synthetic CMD method already applied in many recent works
(see e.g. Cignoni et al. 2012 and references therein). The CMDs of the two observed fields
have been compared with a library of synthetic CMDs computed with
different values of metallicity, ages, and reddening. The synthetic CMDs have
been calculated using evolutionary tracks by Marigo et al. (2008) for masses
between the hydrogen-burning limit (at 0.1 $M_{\odot}$) and $100~M_{\odot}$. Theoretical
temperature and luminosity have been transformed into the observational plane
in the ACS/HRC VEGAMAG photometric system using the trasformations by Girardi et
al. (2008).

The recipe to build a "synthetic" CMD is the following. Using a Monte
Carlo algorithm, masses and ages are extracted according to the initial mass
function (IMF)
and the SF law. The extracted synthetic stars are placed in the CMD by
interpolation among the adopted stellar evolution library. The synthetic
population is put at the distance of the region we want to analyse,
simultaneously correcting for reddening and extinction. Then, to each
synthetic star an artificial star with similar colours and magnitude is 
associated (see Sect. \ref{obs_sec}), and
its output-input magnitudes differences are added to the magnitudes of the 
synthetic star. This last step simultaneously accounts for
photometric errors, incompleteness and blending.

In order to reduce the computational time, a generic SFH is built up
from a linear combination of simple synthetic stellar populations. The
Star Formation Rate (SFR) is parameterized as a linear combination of
chosen basis CMDs, where each basis is a Monte Carlo extraction from a
step star formation episode. The basis CMDs are constructed with all the
metallicities available in the adopted stellar library. 
No metallicity is assumed a priori.

Within the framework of the adopted stellar tracks and atmosphere
models, we chose as the most likely SFH the one which minimizes the differences
between data and synthetic star counts over CMD boxes $0.1$ mag wide
in the V band and $0.05$ mag wide in colour. The degree of likelihood
is assessed through a $\chi^{2}$ minimization and a downhill simplex
algorithm.  In order to escape from local minima, the simplex is
re-started from thousands of initial random positions and a
temperature parameter is implemented. A bootstrap method is used to
assess the effect of random errors. The search of the best SFH is
repeated for each bootstrapped data set, producing a distribution of
best solutions. The error bars on the final SFH represent one standard
deviation using 100 bootstraps.

The SFH of the {\it central field} has been recovered 
both for the F555W vs. (F555W-F814W) and the F555W vs. (F330W-F555W) CMDs
independently while for the {\it offset field} (dominated by red
stars; see Sect. \ref{cmd_sec}) only the F555W vs. (F555W-F814W) CMD has
been used.

We excluded from the analysis the regions containing the known YMCs and
associations up
to the distance where the star density is equal to that of the
surrounding population (see Sect. \ref{age_sec}). We adopted a fixed
distance modulus of $(m-M)_{0}=27.41$ (Dalcanton et al. 2009) and the
IMF of Kroupa (2001). Although recent
works suggest the occurrence of significant variations of the IMF shape 
in different galaxy types (Hoversten \& Glazebrook 2008; Meurer et al. 2009; Lee et al. 2009; 
Gunawardhana et al. 2011), the analysis of this
effect is beyond the aim of our work, so we keep the IMF shape fixed. 
As the distribution of dust in NGC 4214 has been
found to be patchy ({\'U}beda et al. 2007b), we considered the
possibility of differential reddening across the field. For this
purpose, we adopt a gaussian distribution of E(B-V) with central value
and dispersion chosen to best reproduce the location of bright
($F555W<22$) stars in the (F330W-F555W) vs. (F555W-F814W) colour-colour
diagram. In this magnitude range only very young stars ($t<10$ Myr)
are present, so the age-reddening degeneracy is minimized. The best fit
reddening for the {\it central field} turns out to be centered at
E(B-V)=0.2 with a standard deviation $\sigma_{E(B-V)}=0.1$.  Instead,
the {\it offset field} appears to be characterized by a single low
value of E(B-V)=0.05, which is compatible with the foreground
reddening predicted by Schlegel, Finkbeiner \& Davis (1996) maps.

Assuming these reddening values, we searched for the best fitting SFH
by changing the metallicity. We found that Z=0.008 is the best
compromise to reproduce at the same time the BP and the RSG sequence:
higher metallicities tend to split the BP into two separate sequences
(MS and blue HeB stars), while lower values tend to produce a too blue
RSG sequence. This value is compatible with the range of
metallicities ($-0.5<[M/H]<0$) found by W11 by fitting the young
population and is slightly higher than that found by Kobulnicky \& Skillman (1996) using
nebular abundances ($[O/H]\sim-0.5$). Fig. \ref{sfh_field} shows the derived SFH for the
two fields, while Figs.  \ref{cmd_field1} and \ref{cmd_field2} show
the observed and the best fit synthetic CMDs respectively.

The observational and
synthetic CMDs agree with each other. In the central field the BP is well matched in
both the F555W vs. (F330W-F555W) and F555W vs. (F555W-F814W) planes, except above
$F555W=20$, where both the synthetic CMDs are less populated than the
observational ones. Some differences appear also in the RSG phase. In the
F555W vs. (F555W-F814W) plane the upper part of the synthetic RSG is
bluer and shorter, thus suggesting that the red edge of the blue loop
phase in the model is too blue. On the other hand, in the
F555W vs. (F330W-F555W) plane our best model produces a clear sequence of HeB
stars (HeB stars at the blue edge of the blue loop), while the
observed distribution is smooth. This
continuity between MS and HeB stars is puzzling, since Marigo et
al. (2008) isochrones (see Fig. \ref{cmdiso}) predict a clear gap (the so-called
Blue Hertzsprung gap, which is predicted by many models; see
e.g. Chiosi 1998) between the blue edge of the HeB and the
MS. This would imply that the stellar models
predict longer blue loops at these metallicities, a circumstance that could
be favored e.g. by a significant overshooting from convective envelopes (Bressan,
private communication).

In the offset field the agreement is generally better. Our synthetic BP
and RSG sequences reproduce well star counts and morphologies of the
observational counterparts although a mild
colour split along the synthetic BP and a slightly larger ratio of HeB/RSG
stars are visible.

We interpret the differences between observational and theoretical CMDs as being, at least in part, an effect
of the adopted stellar tracks. A likely explanation for the lack of
bright BP stars in the models is that several BP stars are not MS
stars but actually intermediate/massive HeB stars, which are predicted
to be too red in our model. Indeed, the synthetic F555W vs. (F330W-F555W) CMD
shows a clear separation between MS and HeB stars, which is not seen in
the observations. A similar discrepancy is also found by Larsen et
al. (2011) in YMCs. Although a lower metallicity could
alleviate this issue, increasing the excursion to the blue of the HeB
stars, it would also produce a bluer RSG in the optical CMD, thus
exacerbating the difference between data and model in the
red. Alternative explanations could be the rate of mass loss and
rotation. Indeed, stronger mass loss during the RSG phase could favour
a bluewards evolution (see e.g. Salasnich, Bressan \& Chiosi 1999) and, in turn, a
higher ratio between blue and red supergiants. Moreover, when rotation
is taken into account, a large fraction of the end of the core HeB
phase is spent in the blue. In this respect, the new rotating
models of Ekstr{\"o}m et al. (2012) predict that the ratio between blue and red
lifetime for a 20 $M_{\odot}$ model could be up to 7.5 times higher
than for the not-rotating models of Schaller et al. (1992). A
possible solution could result from the inclusion of significant overshooting from
convective envelopes in the stellar models (Bressan, private communication). 
We then think that stellar
models are the major source of discrepancy.

Because of the relatively bright limiting magnitude of our catalog, we
were able to derive the SFH only to few Gyr in lookback time. The {\it
 central field} appears to be characterized by a double peaked SFH
with a prominent episode of SF at $t\sim300$ Myr and another recent
intense episode of SF at $t\sim8$ Myr. The SFH derived on the basis of
the F555W vs. (F555W-F814W) and the F555W vs. (F330W-F555W) CMDs agree
with each other. Of course, because of the low sensitivity of the
ultraviolet filter F330W to the old red populations, the SFH derived
using the ultraviolet CMD is truncated at a more recent lookback time.

The {\it offset field} shows an almost constant SF
rate in the last few Gyr. Recently, McQuinn et al. (2010) and W11 derived the SFH of different
regions of NGC 4214 using deeper HST data, although with worse spatial resolution. 
The SFH derived by McQuinn et al. (2010) is characterized by two major 
epochs of enhancement. The older
episode took place $\sim$3 Gyr ago and produced most of the stars,
the recent one started 1 Gyr ago and is now at its highest
level. A similar behaviour has been found by W11 in their inner field
(covering a wide area of $\sim123\arcsec \times 136\arcsec$ around the
galaxy center), although the older episode has been estimated at more remote
epochs ($\sim$8 Gyr).   
Compared to our solution, intermediate age and recent rates agree
well. In particular, we also predict a low-activity period until 1 Gyr ago, 
followed by a growing activity up to now. There
are differences in the positions of the secondary peaks, but it is not
surprising since our observations allow a higher resolution and cover
the innermost portion of the NGC 4214 active region where the recent
episodes of SF have occurred. Concerning the early activity,
the limiting magnitude of our observations prevents us from reaching
populations older than few Gyr, therefore missing the signature of
the prominent ``old" population detected by W11. Finally, 
the SFH derived in this work
for the old ($>200~Myr$) stellar population is completely reliant on AGB stars. The
modelling of this evolutionary phase is largely uncertain and can produce
significant shifts in the derived ages (Melbourne et al. 2012).

\subsection{Spatial distribution}
\label{spat_sec}

\begin{figure*}
 \includegraphics[width=12cm]{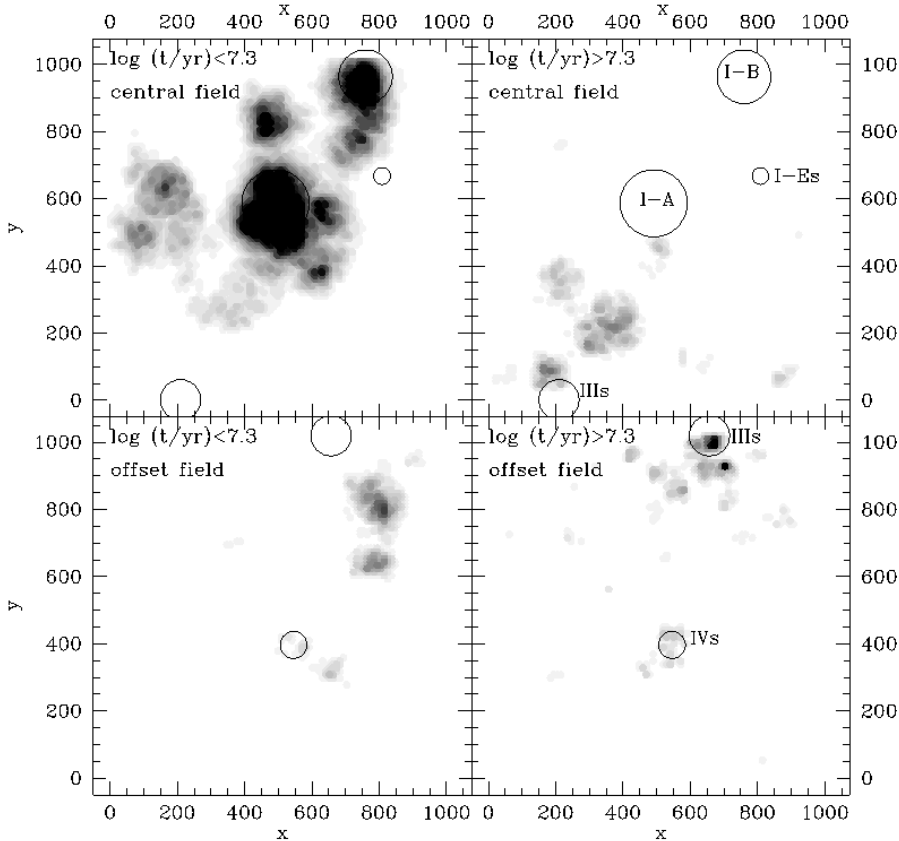}
\caption{Spatial distribution of the young ($\log{t/yr}<7.3$; left panels) and
old ($\log{t/yr}>7.3$; right panels)
populations in the two observed fields of NGC 4214. Top panels refers to the
{\it central field}, bottom panels to the {\it offset field}. The locations of YMCs
and associations are marked with open circles. Density contours range from 3 to 10 times
density standard deviations above the background level.}
\label{map}
\end{figure*}

The SFHs derived in Sect. \ref{sfh_sec} show a
clear distinction between a recent ($\log{t/yr}<7$) episode and an older
($\log{t/yr}>7.5$) one. To study the spatial distribution of the two
episodes of SF we adopted the matched filter technique (Rockosi et al.
2002). Briefly, for each observed field the best fit synthetic
F555W vs. (F555W-F814W) CMD has been split in two samples according
to the star ages ($\log{t/yr}\lessgtr$7.3). To each observed star, a
weight proportional to the ratio of densities in the CMD of stars
belonging to the young/old synthetic populations, calculated at the
observed star location, has been assigned. The above procedure allows us
to highlight the young (old) population by putting the density of
young (old) stars at the numerator of the density ratio. To evaluate
the star density in the CMD, we adopted a metric based on the
Euclidean distance of each star from the F555W and F814W magnitudes of
its neighbor and calculated the density on the basis of the distance
of the 5th nearest star. The final density map in the x,y plane has
been then calculated by using a fixed gaussian kernel estimator
(Silverman 1986) where the volume of each component has been assigned
proportionally to the star weight. The obtained spatial distribution
of the young and old populations in the two observed fields are shown
in Fig. \ref{map}. A clear correlation is apparent between the location of the
young/old objects with that of the corresponding field populations. In particular, 
the two major associations I-A and I-B dominate
the young population of the {\it central field}, with a possible
evidence of a bridge of stars connecting these two major associations. Other
assemblies of young stars are also evident across the field of
view. The old population appears to be segregated from the young
component, populating mainly the bottom part of the field of view close to
the location of the (relatively old; see Sect. \ref{age_sec}) cluster IIIs. A
notable exception to such a correlation is represented by cluster I-Es: despite
its relatively old age (see Sect. \ref{age_sec}), it is located close to the young
field population. In
the {\it offset field} the young population is almost entirely located
in a large scale diffuse region in the upper part of the field of view
while the old population mainly surrounds the two older YMCs present in the
field. 

\section{Young Massive Clusters}

\subsection{Ages}
\label{age_sec}

\begin{table}
 \centering
 \begin{minipage}{140mm}
  \caption{Adopted apertures and derived parameters of YMCs and associations in NGC 4214.}
  \begin{tabular}{@{}lccccr@{}}
  \hline
   cluster & $R_{ap}$ & $\log{t/yr}$ & $\log{M/M_{\odot}}$ & $r_{h}$   & $r_{h}$\\
           & $\arcsec$        &              &                     & $\arcsec$ & pc\\
 \hline
  I-A  & 2.700  & 6.4	     & 5.02        & 1.58        & 23.2\\       
       & 	& ($\pm$0.1) & ($\pm$0.21) & ($\pm$0.12) & ($\pm$2.0)\\ 
  I-B  & 2.160  & 6.8        & --          & --          & -- \\        
       & 	& ($\pm$0.1) & --          & --          & -- \\        
  I-Es & 0.675  & 7.7        & 4.45        & 0.15        & 2.25\\      
       & 	& ($\pm$0.3) & ($\pm$0.32) & ($\pm$0.01) & ($\pm$0.15)\\ 
  IIIs & 1.620  & 8.1        & --          & --          & --\\         
       & 	& ($\pm$0.1) & --          & --          & --\\         
   IVs & 1.080  & 8.0        & 5.11        & 0.23        & 3.43\\       
       & 	& ($\pm$0.1) & ($\pm$0.33) & ($\pm$0.02) & ($\pm$0.35)\\ 
\hline
\end{tabular}
\end{minipage}
\end{table}

\begin{figure}
 \includegraphics[width=8.7cm]{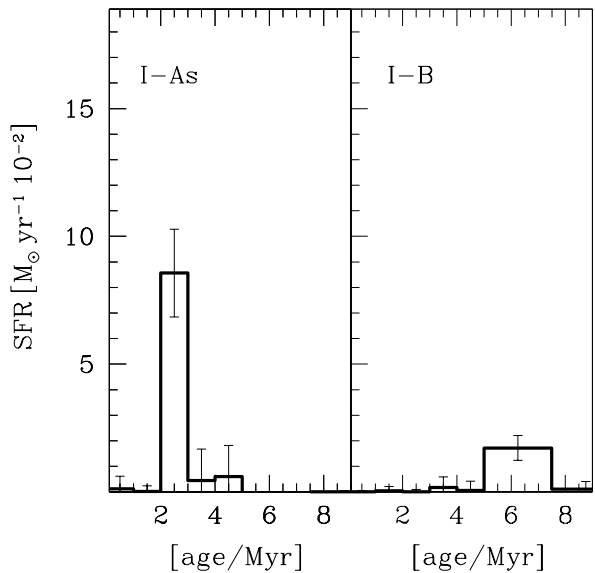}
\caption{SFH derived for associations I-A (left panel) and I-B (right panel).}
\label{ages_sfh}
\end{figure}

\begin{figure*}
 \includegraphics[width=12cm]{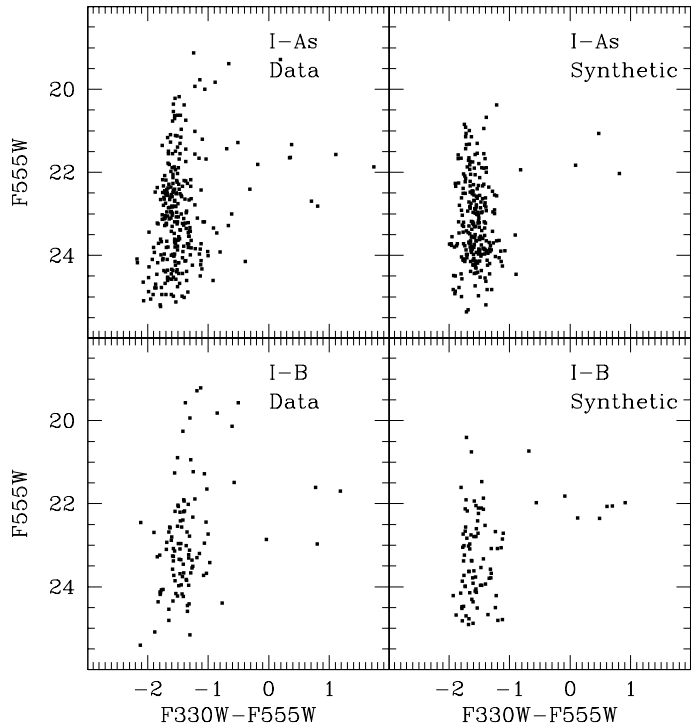}
\caption{Comparison between the observed (left panels) and synthetic (right
panels) F555W vs. (F330W-F555W) CMDs of associations I-A (top panels) and I-B (bottom
panels).}
\label{ages_cmd}
\end{figure*}

\begin{figure*}
 \includegraphics[width=12cm]{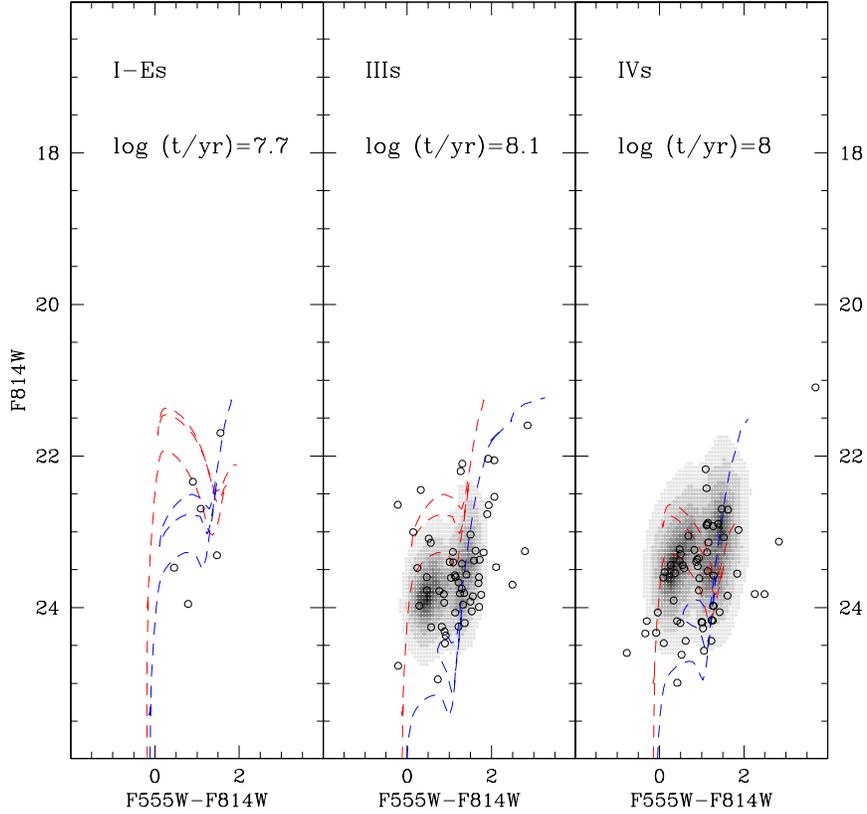}
\caption{Comparison between the observed F814W vs. (F555W-F814W) CMDs of the few 
resolved stars (open circles) of the clusters I-Es, IIIs, and IVs and the best 
fit synthetic models. The isochrones with ages $\Delta~log{t/yr}\pm0.2$ with
respect to the best fit models are marked with dashed lines. 
For clusters IIIs and IVs the Hess diagrams of the best fit models are shown.}
\label{age_all}
\end{figure*}

\begin{figure*}
 \includegraphics[width=12cm]{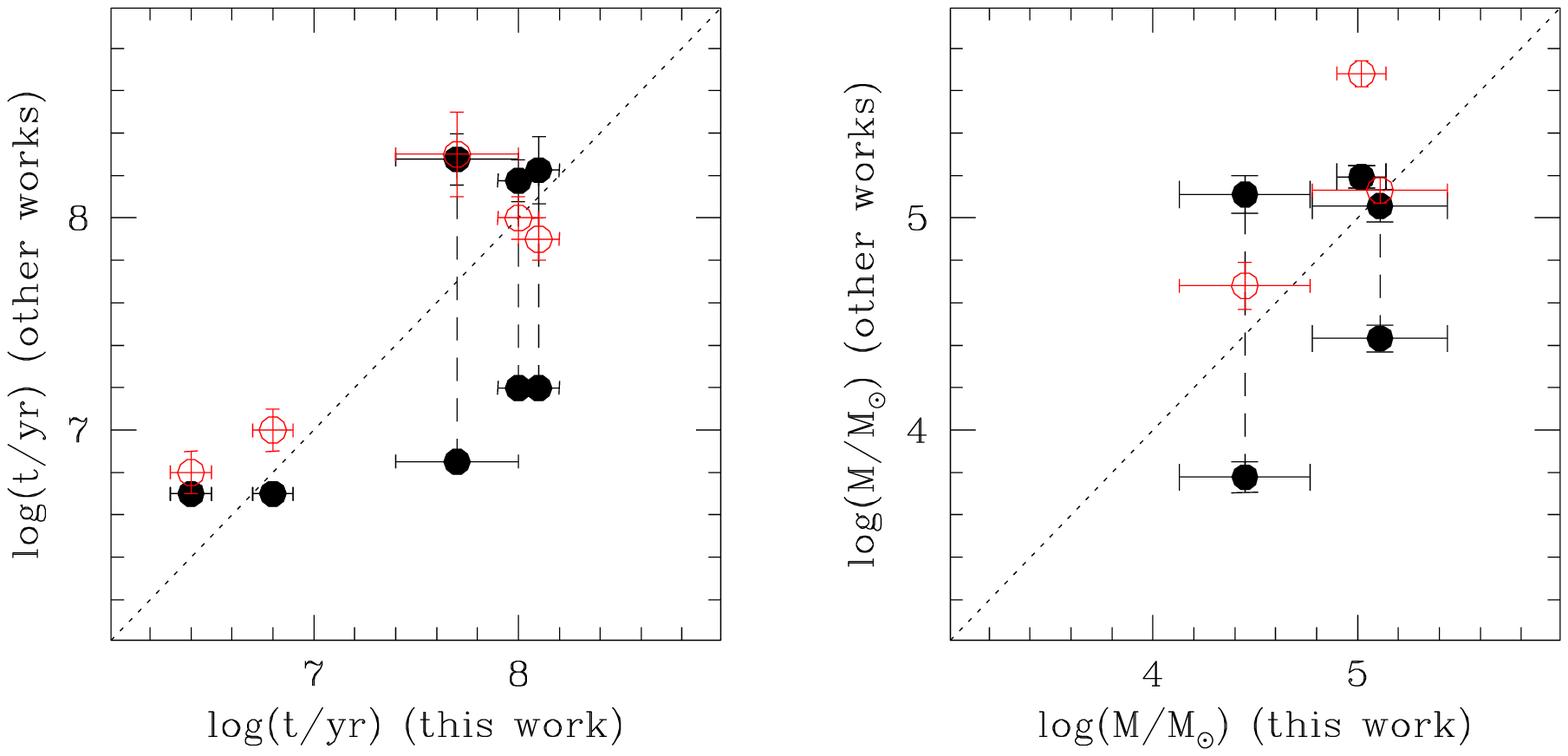}
\caption{Comparison between the ages (left panel) and masses (right panel) 
derived in this work with those by {\'U}beda et al.
(2007b; black dots) and Sollima et al. (2013; open dots). Dashed lines indicate the range covered by the multiple
solutions found by {\'U}beda et al. (2007b). Dotted lines mark the one-to-one
relation.}
\label{confr}
\end{figure*}

The superb resolution of the HRC allows one to resolve 
individual stars in the two major associations (I-A
and I-B) and to partially resolve the outskirts of the other three
analysed clusters (I-Es, IIIs and IVs). This allows a precise age
dating of these clusters as well as an estimate of their sizes and
masses. The ages of the two resolved associations have been derived adopting
the same technique described in Sect. \ref{sfh_sec}. We considered in
our analysis only (real and artificial) stars within a radius corresponding to the distance from the
cluster center of the locus where the stellar density reaches the level of the
background field. According to this definition, a number of field
contaminants could be present within the aperture. 
Since the artificial stars have been distributed uniformly across the
field, the estimated completeness and blending fractions represent an average
within the adopted aperture. In principle, the different distribution of real
and artificial stars could cause an overestimation of the completeness level and
an underestimate of the blending fraction with some consequent bias in the SFH
and mass determination for this stellar system. However, this affects 
almost exclusively the faint part of the completeness curve, while 
bright ($F555W<22$) stars are almost always recovered even in the most crowded region of the
cluster. Since the age determination in young stellar populations, like those
of associations I-A and I-B, is mainly driven by the bright stars (at
$F555W<21$, where the completeness is $\sim90\%$), no significant bias is expected to affect our determinations. 
To test this last conclusion, for both associations, we repeated the analysis 
in two regions at different distances to the cluster center ($\lessgtr 0.5~R_{ap}$). 
As expected, the bestfit ages for both clusters remain unchanged
regardless of the considered radial sample.
The derived SFHs of
both objects are shown in Fig. \ref{ages_sfh} and the corresponding
observed and synthetic CMDs are plotted in Fig. \ref{ages_cmd}. The
adopted aperture radii ($R_{ap}$) for the five clusters as well as
their derived parameters are listed in Table 1. It is apparent that
both associations are characterized by a single prominent peak with a
small residual older population. The oldest stars (noticeable in the
F555W vs. (F555W-F814W) CMD as a sequence of RSG) are homogeneously
distributed within the aperture, at variance with the blue MS stars. Both
the age and the density of these older stars agree with those of the
surrounding field population. Both associations appear to be very young
($\log{t/yr}<7$) in agreement with previous studies ({\'U}beda et
al. 2007b; Sollima et al. 2013; Andrews et al. 2013). 
No signs of multiple stellar
populations are noticeable in either objects.

A bunch of very bright objects ($F555W<20$), not reproduced by the
synthetic CMDs, can be noticed in both systems. The location of these stars in the CMD is not compatible with any stellar track
with $M<100~M_{\odot}$.The fraction of these
objects in the field is significantly smaller (see Sect. \ref{sfh_sec}).
It is therefore likely that they arise from multiple
blends of bright stars which are favored by the extreme crowding conditions met 
in clusters.

For the other three clusters only a small number of stars in their peripheral
region have been resolved. In these cases, the CMD synthesis technique 
cannot be applied. To determine the age of these YMCs we compare the
location of observed stars with the density of stars predicted by synthetic
CMDs, after correcting for field contamination. For this purpose we adopted the
following procedure:
 
\begin{itemize} 
\item{We simulated a single episode of SF with a large ($>10^6$)
    number of stars using the stellar models and the prescription on
    IMF, metallicity, reddening and distance reported in
    Sect. \ref{sfh_sec} and different ages. The synthetic CMD has been
    corrected for completeness and photometric errors using the set of
    artificial stars described in Sect. \ref{obs_sec}.}
\item{We selected the CMD of the reference field population extracted
    in an annulus covering $\sim$4 times the cluster area.}
\item{We calculated the density of stars belonging to the cluster
    catalog ($\rho_{cl}$), the field catalog ($\rho_{field}$) and the
    synthetic catalog ($\rho_{synth}$) at the position of each
    observed object in the F814W vs. (F555W-F814W) CMD. For this
    purpose, we adopted a metric based on the Euclidean distance of
    each synthetic star from the observed F555W and F814W magnitudes
    and calculated the density on the basis of the distance of the 5th
    nearest star. The density of field stars has been normalized to
    the ratio of the field and cluster areas.}
\item{We calculated the merit function
$$L=\frac{\sum_{i=1}^{N_{cl}} w_{i}~\rho_{synth}}{N_{synth}}$$
where $N_{cl}$ and $N_{synth}$ are the number of stars in the observed and
synthetic catalog and
$$w_{i}=1-\rho_{field}/\rho_{cl}$$
is a weight proportional to the probability of a star to be a cluster member, on
the basis of the ratio of densities of field and cluster in the CMD.}
\end{itemize} 
The above procedure has been repeated adopting various ages and the
value maximizing the above defined merit function has been selected as
the best fit. To evaluate errors on the derived ages, the above
procedure has been repeated 100 times by replacing the observed
catalog with other realizations of synthetic CMDs with the same number
of stars $N_{cl}$. The standard deviations of the derived ages has
been assumed as the formal 1 $\sigma$ errors. In Fig. \ref{age_all}
the F555W vs. (F555W-F814W) CMD of these three clusters are compared
with the Hess diagrams of the best fit synthetic populations. It is
noticeable that these three YMCs are all significantly older than the
two major associations (with ages $\log{t/yr}>7.5$), although the
sparse number of stars in cluster I-Es makes this age estimate largely
uncertain. In Fig. \ref{confr} the ages derived here are compared
with those estimated by {\'U}beda et al. (2007b) and Sollima et
al. (2013) on the basis of their integrated photometry. The agreement
is good, without significant offsets and a r.m.s. of
$\sigma_{\log{t/yr}}=0.75$ (with respect to {\'U}beda et al. 2007b)
and $\sigma_{\log{t/yr}}=0.39$ (with respect to Sollima et al. 2013),
compatible with the combined errors of both studies. In this regard,
it is interesting to note that for the three YMCs with multiple age
solutions in {\'U}beda et al. (2007b) (I-Es, IIIs and IVs) our
resolved photometry favors the older solution. Andrews et al. (2013)
provided ages for the young associations of NGC 4214 obtaining for I-A and I-B
ages of 4.2$\pm$1.6 and 2.6$\pm$0.5 Myr (4.8 Myr for both clusters adopting
different SED prescriptions), respectively. The ages estimated by
these authors are generally similar to those found in this paper, and the
small differences are likely due to the different models and  
the lower reddening adopted by these authors.

\subsection{Half-light radii}
\label{rad_sec}

\begin{figure}
 \includegraphics[width=8.7cm]{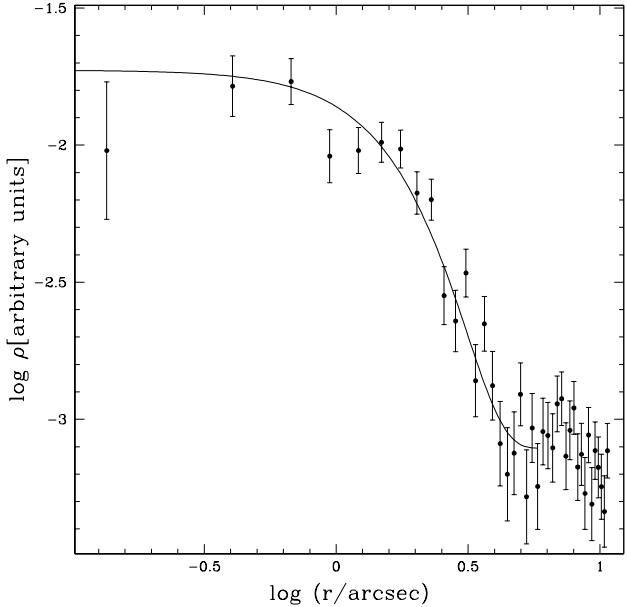}
\caption{Number density profile of association I-A. The best fit King (1966) model
is overplotted.}
\label{prof}
\end{figure}

Another important information we can derive from our photometry are the
half-light radii of the observed stellar complexes. For the association I-A, where we could
resolve the innermost region, this has been done by counting the
number of stars in concentric annuli, correcting their number for the
local completeness level and dividing by the annulus
area. Unfortunately, the center of association I-B lies outside the edge
of the detector in the {\it central field}, so we could not determine
its radial profile. The center of association I-A has been derived
through an iterative procedure in which at each iteration the x and y coordinates of
stars within circles of 100 px radius were averaged and the resulting mean value was
then adopted in the next iteration as the candidate center.
We selected only stars brighter than $F555W<22$ (where the
completeness level is $\sim50\%$ in the innermost region) to avoid
large completeness corrections. The radial density profile of association
I-A is shown in Fig. \ref{prof}. A King (1966) model has been fitted
to the obtained profile to estimate the object half-light radius.

For the marginally resolved clusters we determined their half-light radii
by bestfitting the 2D F555W flux distribution with a set of King (1962) models 
convolved with the HRC Point Spread Function
(Krist, Hook \& Stoher 2010). For this purpose we used ISHAPE 
(Larsen 1999), a specially designed software to derive the structural 
parameters of slightly resolved astronomical objects. With this procedure we
were able to derive the half-light radii of clusters I-Es and IVs, while cluster
IIIs (located at the edge of the detector field of view) has been excluded from
the analysis. 

The half-light radii of the three analysed stellar systems are reported in
Table 1.

\subsection{Masses}
\label{mass_sec}

\begin{figure*}
 \includegraphics[width=12cm]{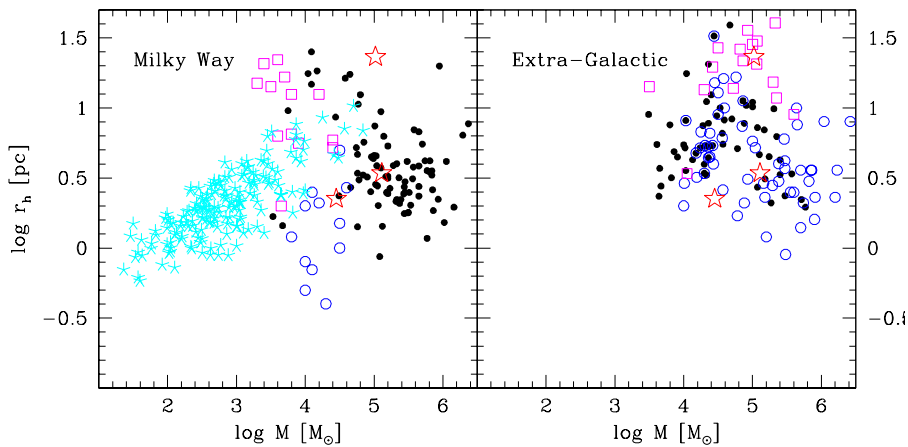}
\caption{Mass vs. half-light radius plane for open clusters (cyan asterisks),
compact YMCs (blue open dots), associations (magenta squares) and GCs
(black filled dots). The left panel refers to Milky Way objects, the right panel refers
to extra-Galactic objects. The location of the three YMCs of NGC 4214 for which masses
and half-light radii have been estimated here are marked in both panels with red
open stars symbols.}
\label{fp}
\end{figure*}

Masses have been determined for the stellar systems entirely sampled in the
two pointings (I-A, I-Es and IVs). For the association I-A, where we could
apply the CMD synthesis technique, the mass is constrained by matching
the number of observed stars. As the different radial distribution of
artificial and real stars could produce significant bias in the completeness 
estimate at faint magnitudes (see Sect. \ref{age_sec}), we corrected the 
observed number of stars in the bright portion of the CMD (at $F330W<22.4$) 
adopting a local completeness estimate: the completeness factor of each star 
has been estimated using the artificial stars with distance from the cluter center
and magnitudes within 10 px and 0.25 mag from those of the considered star.
The contribution of the field population has been estimated in an annulus
surrounding the cluster and removed. 
Moreover, since the limiting magnitude of our
photometry samples only relatively massive ($M>3~M_{\odot}$) stars, an
extrapolation has been done at lower masses assuming a Kroupa (2001)
IMF. For the other two clusters (I-Es and IVs) masses have been
determined from their integrated F555W magnitude. For this purpose, we
used the PHOT task of the DAOPHOT package (Stetson 1987). We chose a
large aperture covering the entire cluster extension and a surrounding
annulus to subtract the background\footnote{To properly estimate the background
we calculated the mean of the sky pixel (instead of the mode
commonly adopted by the PHOT/DAOPHOT algorithm) to account also for the contribution
of the resolved background stars.}. F555W luminosities have been
therefore transformed into masses by adopting the $M/L_{F555W}$ ratio
appropriate for their best fit age (calculated by folding the selected
isochrone with a Kroupa 2001 IMF). For these two older clusters a mass
contribution from white dwarfs has been added assuming the
initial-final mass relation by Kruijssen (2009). The use of the integrated
magnitude to estimate the clusters' masses could lead to potentially important
biases linked to the uncertainties of $M/L_{F555W}$ ratios and the stochastic
errors due to the contribution of few bright stars (Fouesneau et al. 2010;
Popescu \& Hanson 2010). Note however that the magnitude of stochastic effects
drastically decreases at ages $\log{t/yr}>8$ and masses $\log{M/M_{\odot}}>4$
(Beerman et al. 2012). For the typical ages and masses of our targets this
translates into a systematic uncertainty of 
$\sigma_{\log{M/M_{\odot}}}\sim0.2-0.3$. 
The obtained masses
are reported in Table 1. A comparison with the masses estimated by {\'U}beda et
al. 2007b) and Sollima et al. (2013) is shown in the right panel of
Fig. \ref{confr}. Also in this case, there is a good agreement between
all measures without any systematical difference and a r.m.s. of
$\sigma_{\log{M/M_{\odot}}}=0.47$ (with respect to {\'U}beda et
al. 2007b) and $\sigma_{\log{M/M_{\odot}}}=0.49$ (with respect to
Sollima et al. 2013), compatible with the combined uncertainties of
those works. To classify the above objects according to the definition
proposed by Billet et al. (2002), we need to convert their unit
(integrated $M_{V}$ magnitudes at 10 Myr) into solar masses. By
adopting the stellar models, metallicity and IMF reported in
Sect. \ref{sfh_sec}, we derived a limiting mass for a super star
cluster of $\log{M/M_{\odot}}>4.5$ and $4.1<\log{M/M_{\odot}}<4.5$ for
a popoulous cluster. According to this criterion, clusters I-A and IVs
can be classified as super star clusters, while cluster I-Es as a
populous cluster. The same classification has been given by Billet et
al. (2002) for clusters I-Es and IVs, while the association I-A was not included in their
sample.

In Fig. \ref{fp} the location of the three YMCs in the mass
vs. half-light radius plane is compared with those of YMCs,
associations (from Portegies-Zwart et al. 2010), open clusters (from
Piskunov et al. 2007) and GCs (from McLaughlin \& van der Marel 2005)
of the Milky Way and other galaxies. The distinction between YMCs and
associations has been made on the basis of the ratio between age and
dynamical time ($t_{age}/t_{dyn}\lessgtr 3$; Portegies-Zwart et
al. 2010). While I-A clearly falls in the region where most
extra-Galactic associations are located (in agreement with its ratio
$t_{age}/t_{dyn}=0.15$), clusters I-Es and IVs lie on the locus of
YMCs. It is interesting to note that cluster IVs lies in a region where
extra-Galactic YMCs and present-day Galactic GCs overlap.

It is interesting to analyse the fraction of mass contained in clusters
(cluster formation efficiency; $\Gamma$) at different ages in the two observed 
fields. For this purpose, the masses of
the association I-B and cluster IIIs, although only partially sampled by our
observations, have been also estimated adopting the same procedures described
above. In the {\it central field}, $33\pm10$\% of stars younger than 10 Myr are
in one of the two major associations (I-A and I-B). This fraction decreases
to $9\pm3$\% if we consider all stars younger than 200 Myr (including all the 
clusters in this field). A larger fraction has been estimated in the {\it 
offset field} ($19\pm6$\% with $t<200$ Myr). For comparison, the fractions of
stars in clusters at $t<10$ Myr in the samples analysed by Goddard, Bastian \& 
Kennicutt (2010) and Adamo, {\"O}stlin, 
\& Zackrisson et al. (2011) range from 4 to 25\%, with an increasing trend 
with the SFR density.
Considering the mean SFR of the last 10 Myr in the {\it central field}
($\Sigma_{SFR}(<10~Myr)=0.27~M_{\odot}~yr^{-1}~kpc^{-2}$), the fraction of stars in 
clusters estimated here appears to lie slightly above the relation by Goddard et
al. (2012), although within the uncertainties. 
Silva-Villa \& Larsen (2011) and Cook et al.
(2012) considered an age limit of 100 Myr and a sample of galaxies with SFR 
densities which are significantly smaller than the averages in our fields
($\Sigma_{SFR}(<200~Myr)=0.15$ and $0.07~M_{\odot}~yr^{-1}~kpc^{-2}$ for the
{\it central} and the {\it offset} field, respectively). Again, they 
estimated values of $\Gamma$ on average smaller than those found here.
It is worth noting that we calculated $\Gamma$ including stars in
associations which are supposed to dissolve in few Myr. On the other hand,
according to the definition given in Sect. \ref{mass_sec}, the criterion to
distinguish clusters from associations is based on the comparison between age
and dynamical time, this last quantity depending on the system radius. 
In many previous literature works such a cluster/association discrimination 
was not possible because of the missing information about radii. So our estimate
should be comparable with those available in the literature.

\subsection{Mass Function of association I-A}

\begin{figure}
 \includegraphics[width=8.7cm]{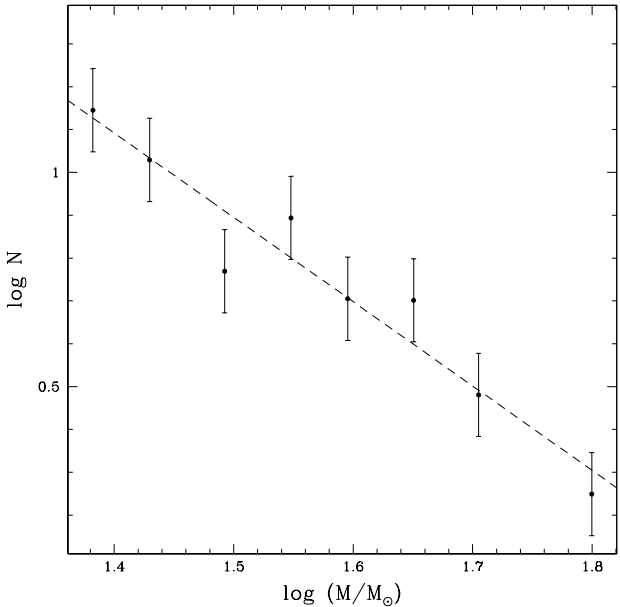}
\caption{MF of association I-A. The best fit power-law (almost corresponding to a
Salpeter 1955 law) is marked with a dashed line.}
\label{mf}
\end{figure}

In the resolved association I-A the mass range where a good 
completeness
level is achieved is relatively large ($27<M/M_{\odot}<72$). It is
therefore possible to estimate the MF of this stellar systems. For this
purpose we converted the F330W magnitude (where the MS is clearly
separated from the other evolutive sequences) into mass by means of the
isochrone which best fits the cluster's CMD. Only MS stars with colours
within $-2<(F330W-F555W)<-1$ and with $F330W<22.4$ have been considered, to
avoid contamination from evolved stars and large completeness
corrections. For reference, in this color range, the completeness $\Psi$ is always 
$>$50\% with $\Psi\sim80\%$ at F330W=21 and $\Psi\sim90\%$ at F330W=20. 
Apparent magnitudes have been transformed into absolute
ones adopting the distance modulus and mean reddening reported in
Sect. \ref{sfh_sec}. To each star a completeness factor has been
assigned according to its distance from the center and to its
magnitude using the neighbor objects (within 10 px) in the artificial
star catalog within 0.25 mag in the F330W and F555W bands. We adopted the method
described by Ma{\'{\i}}z-Apell{\'a}niz \& {\'U}beda (2005) to bin our data
including 20 stars in each bin. The
completeness-corrected MF for association I-A is shown in Fig. \ref{mf}. A
least-square fit of the observed MF yields a slope of
$\alpha=-1.97\pm0.23$, which is slightly flatter than the Salpeter (1955) law.
Consider that the formal uncertainty 
reported here does not account for all the sources of errors
affecting the MF slope determination. A more realistic uncertainty of $\pm 0.4$ 
can be estimated according to the relation provided by Weisz et al. (2013b).
Unfortunately, such a large uncertainty prevents 
any firm conclusion on this issue.

The level of mass segregation within association I-A has been estimated by
comparing the radial distribution of stars with F330W$<$21 with those
in the range $21<F330W<22$ (roughly corresponding to masses
$M>45M_{\odot}$ and $27<M/M_{\odot}<45$). The distance of each star
has been weighted according to the star completeness as estimated
above.  
It is noticeable that massive stars appear to
be more concentrated than less massive ones. However, a
Kolmogorov-Smirnov test gives a probability of 8\% that the two
samples are drawn from the same distribution. So, the above result
appears not to be statistically significant. Note that the mass
segregation timescale of a 56 $M_{\odot}$ star in such a young cluster
is $t_{ms}=0.01~t_{rh}=0.45~Gyr$ (Spitzer 1969; adopting a mean
stellar mass of $<m>=0.57~M_{\odot}$ appropriate for the cluster age
and IMF) i.e. $\sim$180 times longer than the cluster age.

\section{Discussion}

In this work we studied the stellar population of the most active
region of the nearby starburst galaxy NGC 4214 using deep HST images
sampling both the field galactic population and five complexes. The SFHs
of the field population of the two analysed regions are characterized
by the presence of two main bursts occurred within the last 500 Myr: a
recent episode occurred $\sim 8$ Myr ago, mainly in the central region
of the galaxy, and an older one occurred $\sim 300$ Myr ago which is
spread out across a large area. Our analysis does not allow us to
sample the oldest (at ages $>3~Gyr$) galactic population which has
been found to dominate the stellar content of this galaxy in previous
studies (McQuinn et al. 2010; W11). The spatial distribution of the young
($\log{t/yr}<7.3$) and old ($\log{t/yr}>7.3$) populations indicate a
clumpy SF process with many aggregates of stars displaced in different
portions of the field of view according to their ages. The large majority 
of the young ($\log{t/yr}<7.3$) population is located in the innermost region 
of the galaxy within $\sim$200 pc to the galaxy center. The old population is
present in both the analysed fields and does not show any significant spatial
inhomogeneity. Spatial differences in the SFH have been
already noticed in other dwarf irregulars like e.g. the Magellanic Clouds (Weisz
et al. 2013a; Cignoni et al. 2013) and NGC 6822 (Gallart et al. 1996). In the above mentioned
cases, the regions with strong SF activity are those located in the bar or near
the nucleus of these galaxies, well correlated with the HI column density. These
variations are generally interpreted as the signature of a recent migration of
gas towards the galaxy center which simultaneously shut down star formation in the 
outer regions while dramatically increasing the star formation rate in the 
centre. There is a clear correlation between the
spatial distribution of clusters and field populations: by adopting the same
separation criterion between young/old population ($\log{t/yr}\lessgtr$7.3), it
is apparent that the two young associations are located within the area where
the most intense recent SF activity is present. On the other hand, old YMCs are
generally located at large distances from the galactic center 
(with the exception of cluster I-Es), in regions where only old populations are
present. We recall however that the region analysed here covers only the innermost
700 pc of the galaxy and that the SF activity can present a stochastic pattern (Dohm-Palmer
et al. 2002).

In spite of the general agreement between data and model CMDs,
some differences remain. None of Marigo et al. (2008) models
provide an optimal fit for BP and RSG sequences
simultaneously. Moreover, as already found in other investigations
(see e.g. Larsen et al. 2011) of YMCs, models
predict a clear separation between MS and HeB stars in the F555W vs.
(F330W-F555W) CMD, while observations do not show such a
discontinuity.

We derived accurate ages for the five observed assemblies, while masses and
half-light radii have been estimated for three of them. The derived
ages span a wide range between $6.4<\log{t/yr}<8.1$ in agreement with
those predicted by recent analyses based on integrated photometry
({\'U}beda et al. 2007b; Sollima et al. 2013). The youngest associations
(I-A and I-B) appear still surrounded by a shell of gas which is being
probably expelled by the shocks produced by the supernovae II
explosions and/or the wind of very massive stars occurring in these young 
stellar systems.

By comparing the masses of clusters with the SFR of the surrounding field
populations we estimated a cluster formation efficiency of $\Gamma=33\pm10$\% in
the {\it central field} within 10 Myr, which decreases when older ($t<200$ Myr)
populations are considered. The different cluster formation efficiency for the
two age selections is partly explained by the infant mortality of clusters which
are not able to survive to the gas expulsion phase occurring in the first tens
of Myr. This is also supported by the evidence that the cluster formation
efficiency is larger in the {\it offset field}, where the less dense environment
favors the survival of young assemblies. However, the estimated values of $\Gamma$, while rather uncertain,
follow the empirical relation defined by Goddard et al. (2010) as a function of
the SFR density and are larger than those estimated in less active galaxies by
Silva-Villa \& Larsen (2011) and Cook et al. (2012). This evidence supports 
the hypothesis that active galaxies form and retain stars in clusters more
efficiently than those with a small SFR.

We found no clear signatures of multiple stellar populations in the
two resolved associations (I-A and I-B). This is not surprising since,
{\it i)} the number of resolved stars and the uncertainties on the mass
function hamper the detection of such subtle features, and {\it ii)} 
according to the main proposed scenarios, the second generation of
stars should form only after 10-30 Myr after the evolution of rotating
massive stars (Decressin et al. 2007) or intermediate mass AGB stars
(Ventura et al. 2001).

At face value the mass function of the largest association (I-A) is
flatter than the Salpeter (1955) law. 
A comparison with the compilation by Weisz et al. (2013b) indicates that the
slope derived here for this association lies at the bottom envelope of the 
distribution of their sample (see their Fig. 1). Unfortunately, the intrinsic
scatter of the distribution shown by these authors and the large
uncertainty of our estimate prevent any conclusion on a possible deviation
from a universal Salpeter-like IMF.
A hint of mass segregation has been
found in this stellar system, although the statistical significance of this
evidence is only at $\sim2\sigma$. The dynamical mass-segregation
timescale for the considered mass range is more than 2 orders of
magnitude longer than the cluster age. So, if confirmed, this evidence
would indicate a primordial mass segregation driven by a preferential
formation of massive stars in high-density environments (Klessen
2001). Similar evidences have been already found in Galactic
associations like the Arches cluster (Stolte et al. 2002), the Orion
Nebula Cluster (Hillenbrand \& Hartmann 1998), Westerlund 1 (Brandner
et al. 2008) and NGC 3603 (Harayama, Eisenhauer \& Martins 2008) as
well as in a few extra-Galactic YMCs (de Grijs et al. 2002; McCrady,
Graham \& Vacca 2005; Larsen et al. 2008).

The structural properties of the three objects for which we were able to
determine their masses and half-light radii have been compared with
those of other low-mass stellar systems. While I-A presents the
typical characteristics of large associations, the other two clusters
(I-Es and IVs) have masses and half-light radii compatible with those
of present-day Galactic GCs and extra-Galactic YMCs. For comparison,
YMCs in this mass range have been also observed in nearby starburst
galaxies like NGC 1569 (Hunter et al. 2000; Origlia et al. 2001;
Larsen et al. 2011), NGC 1705 (Annibali et al. 2009; Larsen et
al. 2011), NGC 4449 (Annibali et al. 2011), NGC 1313, NGC 5236 and NGC
7793 (Larsen et al. 2011). The structural evolution of these clusters
deserves particular attention. The two youngest associations (I-A and I-B)
are still in the phase during which gas expulsion due to the
explosions of the first supernovae II is ongoing. This can be deduced
by the presence of the shell-shaped gas emission surrounding
these stellar systems. A comparison of the dynamical timescale of
the association I-A (estimated through the mass and radii derived in
Sect. \ref{mass_sec} and \ref{rad_sec} and eq. 11 of Portegies-Zwart
et al. 2010) with its age indicates a ratio $t_{age}/t_{dyn}=0.13$
indicating that it could still not have reached virial
equilibrium. Unfortunately, estimates of its velocity dispersion are
not available, so it is still not possible to assess the dynamical
status of this stellar system. As a consequence of the ongoing gas expulsion
and the future mass-loss due to stellar evolution, these young
clusters will expand being subject to strong tidally induced
mass-loss. The fate of these objects depends on the amount of residual
gas still present within the cluster and the strength of the galactic
tidal field. For these reasons, it is not clear if they could survive
as star clusters for a long time. A different case is that of
the older, compact YMCs (I-Es, IIIs and IVs): these objects survived
to their initial phase of gas expulsion and will experience only a
minor ($<10\%$) stellar evolution-induced mass-loss. The half-mass
relaxation times of clusters I-Es and IVs (calculated using the
Spitzer 1969 formula and masses and effective radii listed in Table 1)
are $t_{rh}=0.46~Gyr$ (I-Es) and $t_{rh}=1.63~Gyr$ (IVs) i.e.
comparable to those of present-day GCs (McLaughlin \& van der Marel
2005). The evaporation timescale for these clusters ($t_{ev}\sim
140~t_{rh}$; Spitzer 1940) turns out to be larger than the Hubble
time. N-body simulations by Madrid, Hurley \& Sippel (2012) indicate
that the future structural evolution of these stellar systems will be
determined by the strength of the tidal field which drives both the
fraction of mass-loss and the evolution of the half-mass radius. In
this regard, it is important to note that both clusters have large
central concentrations, which make them more resistant to tidal
effects. Cluster I-Es is located close to the center of NGC 4214 and
has a relatively small mass, so it is unlikely that it could maintain
a GC-like mass. On the other hand, cluster IVs is more distant from the
galactic center and has a mass and size similar to those of
present-day Galactic GCs (see Fig. \ref{fp}). Under these conditions,
its mass and radius are expected to be only marginally affected by the
weak tidal field of NGC 4214 and could therefore evolve toward a
GC-like structure.

\section*{Acknowledgments}

AB, RG and SL acknowledge the PRIN INAF 2009 "Formation and Early
Evolution of Massive star Clusters" (PI R. Gratton); AS, SL
acknowledge the PRIN INAF 2011 "Multiple populations in globular
clusters: their role in the Galaxy assembly" (PI E. Carretta); and MT,
AB, and SL aknowledge the PRIN MIUR 2010-2011 ``The Chemical
and Dynamical Evolution of the Milky Way and Local Group Galaxies'' (PI
F. Matteucci), prot. 2010LY5N2T. We thank the anonymous referee for his/her
helpful comments and suggestions.


\label{lastpage}


\begin{thebibliography}{99}

\bibitem[Adamo, {\"O}stlin, 
\& Zackrisson 2011]{2011MNRAS.417.1904A} Adamo A., {\"O}stlin G., Zackrisson E., 2011, MNRAS, 417, 1904 
\bibitem[Aloisi, Tosi, 
\& Greggio 1999]{1999AJ....118..302A} Aloisi A., Tosi M., Greggio L., 1999, AJ, 118, 302 
\bibitem[Andrews et 
al. 2013]{2013ApJ...767...51A} Andrews J.~E., et al., 2013, ApJ, 767, 51 
\bibitem[Annibali et al.(2009)]{2009AJ....138..169A} Annibali F., Tosi 
M., Monelli M., Sirianni M., Montegriffo P., Aloisi A., Greggio L., 2009, AJ, 138, 169 
\bibitem[Annibali et al.(2011)]{2011AJ....142..129A} Annibali F., Tosi 
M., Aloisi A., van der Marel R.~P., 2011, AJ, 142, 129 
\bibitem[Arp 
\& Sandage(1985)]{1985AJ.....90.1163A} Arp H., Sandage A., 1985, AJ, 90, 1163 
\bibitem[Beck et al.(2000)]{2000AJ....120..244B} Beck S.~C., Turner 
J.~L., Kovo O., 2000, AJ, 120, 244 
\bibitem[Beerman et 
al. 2012]{2012ApJ...760..104B} Beerman L.~C., et al., 2012, ApJ, 760, 104 
\bibitem[Billett et al.(2002)]{2002AJ....123.1454B} Billett O.~H., Hunter 
D.~A., Elmegreen B.~G., 2002, AJ, 123, 1454 
\bibitem[Brandner et 
al.(2008)]{2008A&A...478..137B} Brandner W., Clark J.~S., Stolte A., Waters R.,
Negueruela I, Goodwin S. P., 2008, A\&A, 478, 137 
\bibitem[Chiosi(1998)]{chiosi98} Chiosi C., 1998, Stellar 
astrophysics for the local group: VIII Canary Islands Winter School of 
Astrophysics, 1 
\bibitem[Cignoni et al.(2012)]{2012ApJ...754..130C} Cignoni M., Cole 
A.~A., Tosi M., Gallagher J. S., Sabbi E., Anderson J., Grebel E. K., Nota A., 2012, ApJ, 754, 130 
\bibitem[Cignoni et 
al. 2013]{2013ApJ...775...83C} Cignoni M., Cole A.~A., Tosi M., Gallagher 
J.~S., Sabbi E., Anderson J., Grebel E.~K., Nota A., 2013, ApJ, 775, 83 
\bibitem[Cole et al. 2007]{2007ApJ...659L..17C} 
Cole A.~A., et al., 2007, ApJ, 659, L17 
\bibitem[Cook et al. 2012]{2012ApJ...751..100C} 
Cook D.~O., et al., 2012, ApJ, 751, 100 
\bibitem[Dalcanton et al.(2009)]{2009ApJS..183...67D} Dalcanton J.~J., 
et al., 2009, ApJS, 183, 67 
\bibitem[Davies et al.(2007)]{2007ApJ...671..781D} Davies B., Figer 
D.~F., Kudritzki R.-P., McKenty J., Najarro F., Herrero A., 2007, ApJ, 671, 781 
\bibitem[D'Ercole et al.(2008)]{2008MNRAS.391..825D} D'Ercole A., 
Vesperini E., D'Antona F., McMillan S.~L.~W., Recchi S., 2008, MNRAS, 391, 825 
\bibitem[Decressin et 
al.(2007)]{2007A&A...464.1029D} Decressin T., Meynet G., Charbonnel C., Prantzos N., 
Ekstr{\"o}m S., 2007, A\&A, 464, 1029 
\bibitem[de Grijs et al.(2002)]{2002MNRAS.331..245D} de Grijs R., Gilmore 
G.~F., Johnson R.~A., Mackey A.~D., 2002, MNRAS, 331, 245 
\bibitem[de Vaucouleurs et al.(1991)]{1991rc3..book.....D} de Vaucouleurs 
G., de Vaucouleurs A., Corwin H.~G. Jr., Buta R.~J., Paturel G., Fouqu\`e P., 1991, Third Reference 
Catalogue of Bright Galaxies.~Volume I-III, Springer eds., New York,
NY (USA)
\bibitem[Dohm-Palmer et 
al. 2002]{2002AJ....123..813D} Dohm-Palmer R.~C., Skillman E.~D., Mateo 
M., Saha A., Dolphin A., Tolstoy E., Gallagher J.~S., Cole A.~A., 2002, AJ, 
123, 813 
\bibitem[Ekstr{\"o}m et al.(2012)]{ekstrom12} Ekstr{\"o}m S., et al., 2012, A\&A, 537, A146
\bibitem[Fanelli et al.(1997)]{1997ApJ...481..735F} Fanelli M.~N., et al., 
1997, ApJ, 481, 735 
\bibitem[Figer et al.(1999)]{1999ApJ...525..750F} Figer D.~F., Kim S.~S., 
Morris M., Serabyn E., Rich R. M., McLean I. S., 1999, ApJ, 525, 750 
\bibitem[Figer et al.(2006)]{2006ApJ...643.1166F} Figer, D.~F., MacKenty, 
J.~W., Robberto, M., Smith K., Najarro F., Kudritzki R. P.,Herrero A., 2006, ApJ, 643, 1166 
\bibitem[Fouesneau 
\& Lan{\c c}on 2010]{2010A&A...521A..22F} Fouesneau M., Lan{\c c}on A., 2010, A\&A, 521, A22 
\bibitem[Frenk et al.(1988)]{1988ApJ...327..507F} Frenk C.~S., White 
S.~D.~M., Davis M., Efstathiou G., 1988, ApJ, 327, 507 
\bibitem[Gallart et 
al. 1996]{1996AJ....112.2596G} Gallart C., Aparicio A., Bertelli G., 
Chiosi C., 1996, AJ, 112, 2596 
\bibitem[Garc{\'{\i}}a-Benito 
\& P{\'e}rez-Montero(2012)]{2012MNRAS.423..406G} Garc{\'{\i}}a-Benito R.,
P{\'e}rez-Montero E., 2012, MNRAS, 423, 406 
\bibitem[Girardi et al.(2008)]{2008PASP..120..583G} Girardi L., 
et al., 2008, PASP, 120, 583 
\bibitem[Goddard, Bastian, 
\& Kennicutt 2010]{2010MNRAS.405..857G} Goddard Q.~E., Bastian N., Kennicutt R.~C., 2010, MNRAS, 405, 857 
\bibitem[Greggio et 
al. 1998]{1998ApJ...504..725G} Greggio L., Tosi M., Clampin M., de Marchi 
G., Leitherer C., Nota A., Sirianni M., 1998, ApJ, 504, 725 
\bibitem[Gunawardhana et al.(2011)]{2011MNRAS.415.1647G} Gunawardhana 
M.~L.~P., et al., 2011, MNRAS, 415, 1647 
\bibitem[Harayama et al.(2008)]{2008ApJ...675.1319H} Harayama Y., 
Eisenhauer F., Martins F., 2008, ApJ, 675, 1319 
\bibitem[Hillenbrand 
\& Hartmann(1998)]{1998ApJ...492..540H} Hillenbrand L.~A., Hartmann L.~W.,
1998, ApJ, 492, 540 
\bibitem[Hoversten 
\& Glazebrook(2008)]{2008ApJ...675..163H} Hoversten E.~A., Glazebrook K.,
2008, Apj, 675, 163 
\bibitem[Hunter et al.(2000)]{2000AJ....120.2383H} Hunter D.~A., 
O'Connell R.~W., Gallagher J.~S., 
Smecker-Hane T.~A., 2000, AJ, 120, 2383 
\bibitem[Karachentsev et al.(2004)]{2004AJ....127.2031K} Karachentsev 
I.~D., Karachentseva V.~E., Huchtmeier W.~K., 
Makarov D.~I., 2004, AJ, 127, 2031 
\bibitem[King(1962)]{1962AJ.....67..471K} King I.~R., 1962, AJ, 67, 471 
\bibitem[King(1966)]{1966AJ.....71...64K} King I.~R., 1966, AJ, 71, 64 
\bibitem[Klessen(2001)]{2001ApJ...556..837K} Klessen R.~S., 2001, ApJ, 
556, 837 
\bibitem[Kobulnicky 
\& Skillman(1996)]{kobu96} Kobulnicky H.~A., Skillman E.~D., 1996, ApJ, 471, 211 
\bibitem[Krist et al.(2010)]{2010ascl.soft10057K} Krist J., Hook R., 
Stoehr F., 2010, Astrophysics Source Code Library, 10057 
\bibitem[Kroupa(2001)]{2001MNRAS.322..231K} Kroupa P., 2001, MNRAS, 322, 
231 
\bibitem[Kruijssen(2009)]{2009A&A...507.1409K} Kruijssen J.~M.~D., 2009, A\&A, 507, 1409 
\bibitem[Kruijssen(2012)]{2012MNRAS.426.3008K} Kruijssen J.~M.~D., 2012, 
MNRAS, 426, 3008 
\bibitem[Larsen 
\& Richtler(2000)]{2000A&A...354..836L} Larsen S.~S., Richtler T., 2000, A\&A, 354, 836 
\bibitem[Larsen(1999)]{1999A&AS..139..393L} Larsen S.~S., 1999, A\&AS, 139, 393 
\bibitem[Larsen et al.(2001)]{2001AJ....121.2974L} Larsen S.~S., Brodie 
J.~P., Huchra J.~P., Forbes D.~A., Grillmair C.~J., 2001, AJ, 121, 2974 
\bibitem[Larsen et al.(2008)]{2008MNRAS.383..263L} Larsen S.~S., Origlia 
L., Brodie J., Gallagher J.~S., 2008, MNRAS, 383, 263 
\bibitem[Larsen et 
al.(2011)]{2011A&A...532A.147L} Larsen S.~S.,et al., 2011, A\&A, 532, A147 
\bibitem[Lee et al.(2009)]{2009ApJ...706..599L} Lee J.~C., et al., 2009, Apj, 706, 599 
\bibitem[MacKenty et al.(2000)]{2000AJ....120.3007M} MacKenty J.~W., 
Ma{\'{\i}}z-Apell{\'a}niz J., Pickens C.~E., Norman C.~A., 
Walborn N.~R., 2000, AJ, 120, 3007 
\bibitem[Madrid et al.(2012)]{2012ApJ...756..167M} Madrid J.~P., Hurley 
J.~R., Sippel A.~C., 2012, ApJ, 756, 167 
\bibitem[Ma{\'{\i}}z-Apell{\'a}niz et 
al.(1998)]{1998A&A...329..409M} Ma{\'{\i}}z-Apell{\'a}niz J., Mas-Hesse J.~M., 
Munoz-Tunon C., Vilchez J.~M., Castaneda H.~O., 1998, A\&A, 329, 409 
\bibitem[Ma{\'{\i}}z Apell{\'a}niz 
\& {\'U}beda 2005]{2005ApJ...629..873M} Ma{\'{\i}}z Apell{\'a}niz J., {\'U}beda L., 2005, ApJ, 629, 873 
\bibitem[Marigo et 
al.(2008)]{2008A&A...482..883M} Marigo P., Girardi L., Bressan A.,
Groenwegen M.~A.~T., Silva L., Granato G.~L., 2008, A\&A, 482, 883 
\bibitem[McCrady et al.(2005)]{2005ApJ...621..278M} McCrady N., Graham 
J.~R., Vacca W.~D., 2005, ApJ, 621, 278 
\bibitem[McLaughlin 
\& van der Marel(2005)]{2005ApJS..161..304M} McLaughlin D.~E., van der Marel
R.~P., 2005, ApJS, 161, 304 
\bibitem[McQuinn et 
al. 2011]{2011ApJ...740...48M} McQuinn K.~B.~W., Skillman E.~D., 
Dalcanton J.~J., Dolphin A.~E., Holtzman J., Weisz D.~R., Williams B.~F., 
2011, ApJ, 740, 48 
\bibitem[Melbourne et 
al. 2012]{2012ApJ...748...47M} Melbourne J., et al., 2012, ApJ, 748, 47 
\bibitem[Meurer et al.(2009)]{2009ApJ...695..765M} Meurer G.~R., et al., 
2009, Apj, 695, 765 
\bibitem[Origlia et al.(2001)]{2001AJ....122..815O} Origlia L., Leitherer 
C., Aloisi A., Greggio L., Tosi M., 2001, AJ, 122, 815 
\bibitem[Parmentier et al.(2008)]{2008ApJ...678..347P} Parmentier G., 
Goodwin S.~P., Kroupa P., Baumgardt H., 2008, ApJ, 678, 347 
\bibitem[Perina et 
al.(2010)]{2010A&A...511A..23P} Perina S., et al., 2010, A\&A, 511, A23 
\bibitem[Piskunov et 
al.(2007)]{2007A&A...468..151P} Piskunov A.~E., Schilbach E., Kharchenko
N.~V., R{\"o}ser S., Scholz R.-D., 2007, A\&A, 468, 151 
\bibitem[Popescu 
\& Hanson 2010]{2010ApJ...713L..21P} Popescu B., Hanson M.~M., 2010, ApJ, 713, L21 
\bibitem[Portegies Zwart et 
al.(2010)]{2010ARA&A..48..431P} Portegies Zwart S.~F., McMillan S.~L.~W., 
Gieles M., 2010, ARA\&A, 48, 431 
\bibitem[Rockosi et al.(2002)]{2002AJ....124..349R} Rockosi C.~M., 
et al., 2002, AJ, 124, 349 
\bibitem[sala99]{1999A&A...342..131S} Salasnich B., Bressan A., Chiosi C.,
1999, A\&A, 342, 131 
\bibitem[Salpeter(1955)]{1955ApJ...121..161S} Salpeter E.~E., 1955, ApJ, 
121, 161 
\bibitem[Sandage 
\& Bedke(1994)]{1994cag..book.....S} Sandage A., Bedke J., 1994, The 
Carnegie Atlas of Galaxies.~Volumes I, II., Carnegie Institution of 
Washington Publ., No.~638  
\bibitem[Schaller et 
al.(1992)]{schaller92} Schaller G., Schaerer D., Meynet G., Maeder A.,
1992, A\&AS, 96, 269 
\bibitem[Schlegel et al.(1998)]{1998ApJ...500..525S} Schlegel D.~J., 
Finkbeiner D.~P., Davis M., 1998, ApJ, 500, 525 
\bibitem[Silva-Villa 
\& Larsen(2011)]{2011A&A...529A..25S} Silva-Villa E., Larsen S.~S., 2011,
A\&A, 529, A25 
\bibitem[Silva-Villa 
\& Larsen(2012)]{2012MNRAS.423..213S} Silva-Villa E., Larsen S.~S., 2012, MNRAS, 423, 213 
\bibitem[Silverman(1986)]{1986desd.book.....S} Silverman B.~W., 1986, 
Monographs on Statistics and Applied Probability, London: Chapman and Hall, 
1986  
\bibitem[Sirianni et al.(2005)]{2005PASP..117.1049S} Sirianni M., et al., 2005,
PASP, 117, 1049 
\bibitem[Sollima et 
al. 2013]{2013MNRAS.433.1276S} Sollima A., Gratton R.~G., Carretta E., 
Bragaglia A., Lucatello S., 2013, MNRAS, 433, 1276 
\bibitem[Spitzer(1940)]{1940MNRAS.100..396S} Spitzer L. Jr., 1940, 
MNRAS, 100, 396 
\bibitem[Spitzer(1969)]{1969ApJ...158L.139S} Spitzer L., Jr., 1969, ApJL, 
158, L139 
\bibitem[Stetson(1987)]{1987PASP...99..191S} Stetson P.~B., 1987, PASP, 
99, 191 
\bibitem[Stolte et 
al.(2002)]{2002A&A...394..459S} Stolte A., Grebel E.~K., Brandner W.,
Figer D.~F., 2002, A\&A, 394, 459 
\bibitem[Tolstoy et 
al.(2009)]{2009ARA&A..47..371T} Tolstoy E., Hill V., Tosi M., 2009, ARA\&A, 47, 371 
\bibitem[Tutukov(1978)]{1978A&A....70...57T} Tutukov A.~V., 1978, A\&A, 70, 57 
\bibitem[{\'U}beda et al.(2007)]{2007AJ....133..917U} {\'U}beda L., 
Ma{\'{\i}}z-Apell{\'a}niz J., MacKenty J.~W., 2007b, AJ, 133, 917 
\bibitem[{\'U}beda et al.(2007)]{2007AJ....133..932U} {\'U}beda L., 
Ma{\'{\i}}z-Apell{\'a}niz J., MacKenty J.~W., 2007a, AJ, 133, 932 
\bibitem[Ventura et al.(2001)]{2001ApJ...550L..65V} Ventura P., D'Antona 
F., Mazzitelli I., Gratton R., 2001, ApJL, 550, L65 
\bibitem[Vieira et al.(2013)]{2013Natur.495..344V} Vieira, J.~D., et al., 2013, Nature, 495, 344 
\bibitem[Weisz et al.2011]{2011ApJ...743....8W} 
Weisz D.~R., et al., 2011, ApJ, 743, 8 
\bibitem[Weisz et al. 2013]{2013MNRAS.431..364W} 
Weisz D.~R., Dolphin A.~E., Skillman E.~D., Holtzman J., Dalcanton J.~J., 
Cole A.~A., Neary K., 2013, MNRAS, 431, 364 
\bibitem[Weisz et al. 2013]{2013ApJ...762..123W} 
Weisz D.~R., et al., 2013, ApJ, 762, 123 
\bibitem[White 
\& Rees(1978)]{1978MNRAS.183..341W} White S.~D.~M., Rees M.~J., 1978, MNRAS, 183, 341 
\bibitem[Williams et al.(2011)]{2011ApJ...735...22W} Williams B.~F., 
Dalcanton J.~J., Gilbert K.~M., Seth A. ~C., Weisz D.~R., Skillman E.~D.,
Dolphin A.E., 2011, ApJ, 735, 22 



\end{thebibliography}
\end{document}